\documentclass[submission, Proceedings]{SciPost}

\hypersetup{
    colorlinks,
    linkcolor={red!50!black},
    citecolor={blue!50!black},
    urlcolor={blue!80!black}
}

\usepackage[bitstream-charter]{mathdesign}
\DeclareSymbolFont{usualmathcal}{OMS}{cmsy}{m}{n}
\DeclareSymbolFontAlphabet{\mathcal}{usualmathcal}

\begin{document}

\begin{center}{\Large \textbf{
Exclusive hadronic $\tau$ decays, within $\&$ beyond the Standard Model\\
}}\end{center}

\begin{center}
P. Roig\textsuperscript{1}
\end{center}

\begin{center}
{\bf 1} Departamento de Física, Centro de Investigación y de Estudios Avanzados del IPN, Apdo. Postal
14-740,07000 Ciudad de México, México
\\
* proig@fis.cinvestav.mx
\end{center}

\begin{center}
\today
\end{center}

\definecolor{palegray}{gray}{0.95}
\begin{center}
\colorbox{palegray}{
  \begin{minipage}{0.95\textwidth}
    \begin{center}
    {\it  16th International Workshop on Tau Lepton Physics (TAU2021),}\\
    {\it September 27 – October 1, 2021} \\
    \doi{10.21468/SciPostPhysProc.?}\\
    \end{center}
  \end{minipage}
}
\end{center}

\section*{Abstract}
{\bf
Semileptonic exclusive tau decays in the Standard Model are reviewed. As it is well-known, they are a privileged arena to learn about low-energy hadronization. They also allow for a number of clean new physics tests, which have attracted much attention recently,  probing scales as high as a few TeV. In light of forthcoming Belle-II data, perspectives on this area are bright.
}

\vspace{10pt}
\noindent\rule{\textwidth}{1pt}
\tableofcontents\thispagestyle{fancy}
\noindent\rule{\textwidth}{1pt}
\vspace{10pt}

\section{Introduction}
\label{sec:intro}
\hspace{0.5cm}The $\tau$ is the only lepton massive enough to decay intro hadrons (mesons). As such, it has been (still is and will continue to be) a privileged tool to learn about the hadronization of QCD currents at low energies in a clean environment \cite{Davier:2005xq, Pich:2013lsa}.

In this workshop we have heard several  talks devoted to this subject. These covered the interesting aspects that can be studied in semileptonic tau decays in the Standard Model (SM), aiming as well to uncover possible effects of new physics beyond it. Specifically, lepton universality $^\dagger$ and CKM unitarity tests $^{\dagger *}$, studies of second class currents $^\circ$, $CP$ and $T$ violation $^\square$, hadronic contributions to the muon g-2 $^\%$, as well as multiparticle final-state interactions and determination of resonance pole parameters $^\triangle$ were the focus of these talks~\footnote{See, respectively, Gabriel López Castro ($^\dagger$), Alberto Lusiani ($^*$), Bachir Moussallam ($^\circ$), Zhi-Hui Guo and Frederic Noël ($^\square$), Álex Miranda ($^\%$), Fabian Krinner and Mikhail Mikhasenko  ($^\triangle$) talks and corresponding contributions to these proceedings. These decays were also touched upon in the more general talks given by Bill Marciano, Zbigniew Was, Xiaorong Zhou, Denis Epifanov, Ami Rostomyan, Mogens Dam and Mike Roney and of course in the introductory historical talk given by Toni Pich and in the closing outlook talk by Hasaka-san (see these proceedings).}.

Here, I will first review briefly hadronic tau decays in the SM and then discuss searches for non-standard interactions making use of them~\footnote{A recent discussion of these topics can be found in ref. \cite{Gonzalez-Solis:2021qzp}.}. I will end with some concluding remarks.

\section{Semileptonic $\tau$ decays in the Standard Model}\label{sec:SM}
\hspace{0.5cm}At leading order, $\tau$ decays either leptonically, into $\nu_\tau \overline{\nu_{e/\mu}} (e/\mu)$, or hadronically, into $\nu_\tau \overline{u} D$ ($D=d,s$) with roughly one third into the former and the remaining two thirds into the latter, detected in a variety of meson final states \cite{ParticleDataGroup:2020ssz}. Rich and large datasets accumulated and analyzed at CLEO, LEP, BaBar and Belle (and soon at Belle-II!) have fostered an increasingly precise knowledge of them over the last decades.

The matrix element for semileptonic tau decays, $\tau^-\to\nu_\tau H^-$, reads
\begin{equation}
    \mathcal{M}=\frac{G_F}{\sqrt{2}}V_{uD} \overline{u}_{\nu_\tau} \gamma^\mu (1-\gamma_5) u_\tau \mathcal{H}_\mu\,,
\end{equation}
where $G_F$ is the Fermi constant, $V_{uD}$ is the CKM matrix element ($D=d,s$ for strangeness-conserving/changing transitions), and the lepton current multiplies the hadron vector
\begin{equation}
    \mathcal{H}_\mu=\left\langle H\Big|(\mathcal{V}_\mu-\mathcal{A}_\mu)\mathrm{e}^{i \mathcal{L}_{QCD}}\Big|0\right\rangle,
\end{equation}
which encodes the hadronization process, creating the final-state mesons from the QCD vacuum, via the left-handed charged weak current, in presence of strong interactions. Lorentz and discrete QCD symmetries allow to decompose $\mathcal{H}_\mu$ in terms of a number of allowed tensor structures times the corresponding form factors (FFs), which are scalar functions of the kinematical invariants~\footnote{An alternative, though equivalent, splitting is given using the so-called structure functions \cite{Kuhn:1992nz}.}. Data is, obviously, of uttermost importance in understanding these FFs. Theoretically, we are assisted by chiral symmetry \cite{Weinberg:1978kz,Gasser:1983yg,Gasser:1984gg}, axiomatic quantum field theory properties, parton dynamics \cite{Lepage:1980fj}, ... Dispersive FF  representations are extremely convenient, as they are best suited to fulfill analyticity, unitarity and crossing symmetry.

In the one-meson case, the relevant $\mathcal{H}^\mu$ is ($P=\pi/K$)
\begin{equation}
    \left\langle P^-(p)\big|\overline{D}\gamma^\mu\gamma_5 u\big| 0\right\rangle=-i\sqrt{2}f_P p^\mu\,,
\end{equation}
where the $P$ decay constant, $f_P$, is known from $P^-\to\mu^-\overline{\nu}_\mu$ ($f_\pi\sim92$ MeV with this, chiral, normalization). If there is new physics, however, such experimentally determined $f_P$ would differ from $f_P^{QCD}$, which is obtained from lattice data \cite{FlavourLatticeAveragingGroup:2019iem}. Radiative corrections (including structure-dependent effects) are essential in elucidating this \cite{Guo:2010dv, Guevara:2013wwa, Arroyo-Urena:2021nil}. LEP measurements of this one-meson tau decays branching ratios \cite{OPAL:2000fde, ALEPH:2005qgp} and by BaBar of their normalization with respect to the lepton channels \cite{BaBar:2009lyd} are fundamental for these tests. Belle observed, for the first time,  $\tau^-\to\pi^-e^+e^-\nu_\tau$ recently~\cite{Belle:2019bpr}. Measurements of these and other radiative processes are essential for the corresponding new physics tests and will hopefully be made at Belle-II.

For two-mesons, a convenient decomposition of $\mathcal{H}^\mu$ is
\begin{equation}
   \left\langle P^-(p)P'^0(p_0)\big|\overline{D}\gamma^\mu u\big| 0\right\rangle=C_{PP'^-} \left\lbrace \left( p_- - p_0-\frac{\Delta_ {PP'}}{s}q\right)^\mu F_V^{PP'}(s)+\frac{\Delta_ {PP'}}{s}q^\mu F_S^{PP'}(s)  \right\rbrace \,,
\end{equation}
where $q^\mu=(p_-+p_0)^\mu$, $s=q^2$ and $\Delta_ {PP'}=m_{P^-}^2-m_{P'^0}^2$. The FFs above, $F_{S/V}^{PP'}(s)$ carry spin$^\mathrm{parity}$ $0^+/1^-$ degrees of freedom,  the scalar one being suppressed by the approximate $SU(3)$ flavor symmetry with respect to the usually dominant vector FF. $C_{PP'}$ are defined so that $F_V^{PP'}=1+\mathcal{O}(s)$ at low $s$ ($C_{\pi\pi}=\sqrt{2}$, for instance).

$\tau^-\to\pi^-\pi^0\nu_\tau$ and $\tau^-\to K^- K_S\nu_\tau$ decays admit a joint description in the isospin symmetry limit \cite{Gasser:1984ux,Oller:2000ug, Gonzalez-Solis:2019iod}. In this case, one can apply the standard (see e.g. refs. \cite{Pich:2001pj, GomezDumm:2013sib, Gonzalez-Solis:2019iod}) thrice-subtracted dispersive description of $F_V^{\pi\pi}(s)$ (and neglect $F_S^{\pi\pi}(s)$)
\begin{equation}
F_V^{\pi\pi}(s)=\mathrm{exp}\left[\alpha_1 s+\frac{\alpha_2}{2}s^2+\frac{s^3}{\pi}\int_{4m_\pi^2}^\infty\mathrm{d}s'\frac{\delta^1_1(s')}{(s')^3(s'-s-i0)}\right]\,,
\end{equation}
where the subtraction constants $F_V(0)=1$ and $\left\lbrace\alpha_i\right\rbrace_{i=1,2}$ determine the polynomial expansion in $s$ of this FF close to the origin. According to Watson's final-state interactions theorem \cite{Watson:1952ji}, $\delta^1_1(s) $ equals the $\pi\pi$ scattering phaseshift in the elastic region, which is extremely well-known \cite{Colangelo:2001df, Garcia-Martin:2011iqs, Caprini:2011ky} and includes the dominant effect of the $\rho(770)$ resonance exchange as well as the leading (resummed) chiral logs.  Asymptotically, this phase goes to $\pi$ (and the FF modulus vanishes as $1/s$). Appropriate (smooth) interpolants for $\delta^1_1(s)$ between these two extreme regions will capture the dynamics of the $\rho(1450)$ and $\rho(1700)$ states, enabling the determination of their mass and width pole parameters \cite{Pich:2001pj, GomezDumm:2013sib, Gonzalez-Solis:2019iod} through fits to the data \cite{OPAL:1998rrm, CLEO:1999dln,ALEPH:2005qgp, Belle:2008xpe, BaBar:2018qry}. Di-pion radiative tau decays have been studied in Refs. \cite{Cirigliano:2001er, Cirigliano:2002pv, Flores-Baez:2006yiq, Miranda:2020wdg, GutierrezSantiago:2020bhy} (see also Z. H. Guo and Á. Miranda, these proceedings) and provide an independent evaluation of the leading hadronic vacuum polarization contribution to the muon g-2 \cite{Aoyama:2020ynm},  $a_\mu^{HVP,LO}$.

In the $K\pi$ tau decay modes, the scalar FF is no longer negligible. It can be obtained by means of a coupled-channel ($K(\pi/\eta/\eta')$) strangeness-changing meson-meson scattering analysis \cite{Jamin:2000wn, Jamin:2001zq}. The dispersive vector FF is constructed in analogy to the $\pi\pi$ case \cite{Moussallam:2007qc, Boito:2008fq}, and can benefit from $K_{\ell3}$ decays data \cite{Boito:2010me, Antonelli:2013usa, Bernard:2013jxa}. In the $SU(3)$ flavor symmetry limit, the $K\eta^{(\prime)}$ vector FFs can be related to the $K\pi$ one \cite{Escribano:2013bca} and a joint fit to $\tau^-\to K_S \pi^-\nu_\tau$ \cite{Belle:2007goc} and $\tau^-\to K^- \eta\nu_\tau$ \cite{Belle:2008jjb} data improves particularly the extraction of the $K^*(1410)$ pole parameters \cite{Escribano:2014joa}. $\tau^-\to K^-\pi^0\nu_\tau$ branching fraction was best measured by BaBar \cite{BaBar:2007yir} but the corresponding  spectrum remains unpublished and the $\tau^-\to K^-\eta'\nu_\tau$ decays have not been discovered yet \cite{BaBar:2012zfq}.

The $\pi^-\eta^{(\prime)}$ decay modes are much more difficult to understand. On the one hand, only upper limits exist on these \cite{BaBar:2010bul,BaBar:2012zfq}. On the other,  G-parity suppresses them \footnote{Other competing isospin-violating effects have been addressed in refs. \cite{Guevara:2016trs,Hernandez-Tome:2017pdc}.} and complicates their analysis \cite{Belle-II:2018jsg}. As a result, two dispersive analyses of these decays  \cite{Descotes-Genon:2014tla,Escribano:2016ntp} differ by an order of magnitude (due to the uncertain scalar contribution, see also Refs. \cite{Guo:2012yt, Dudek:2016cru,Guo:2016zep}) in the predicted branching ratio for the $\eta$ channel (see B. Moussallam, these proceedings).

No fully-dispersive treatment of three-meson tau decays \footnote{Four form factors appear in these processes: two of them carrying $1^+$ degrees of freedom, one corresponding to $1^-$ (which is linked to the chiral anomaly) and another one to $0^+$ (that is suppressed in the chiral limit).} exist, in which three-body final-state interactions are accounted for completely. Chiral Lagrangian studies have been done for the $3\pi$ \cite{Colangelo:1996hs, GomezDumm:2003ku, Dumm:2009va}, $KK\pi$ \cite{Dumm:2009kj} and $\eta\pi^-\pi^0$ \cite{GomezDumm:2012dpx} decay modes (which were incorporated into the corresponding version \cite{Shekhovtsova:2012ra, Nugent:2013hxa} of the TAUOLA library \cite{Jadach:1990mz,Jadach:1993hs}, together with the two-meson tau decays, featuring dispersive form factors, discussed before) and for $\tau^-\to (V P)^- \nu_\tau$ decays \cite{Guo:2008sh}, where two/three of the pseudoscalar mesons ($\pi,K$) are close to the on-shell condition for a vector resonance ($\rho,\omega,\phi$). Dispersive analyses based on two-body final-state interactions were completed for the $3\pi$ \cite{JPAC:2018zwp} and $K\pi\pi$ \cite{Moussallam:2007qc} decay modes (see also M. Mikhasenko,  these proceedings). Best measurements of the spectra of these decays still come from ALEPH \cite{ALEPH:1998rgl,ALEPH:1999jxs,ALEPH:1999uux} and CLEO \cite{CLEO:2004hrb}. We hope Belle-II will finally improve upon them. Higher multiplicity modes are omitted in this overview.

\section{Semileptonic $\tau$ decays beyond the Standard Model}
\label{sec:BSM}
\hspace{0.5cm}For new physics searches, the main advantage of the effective field theory (EFT) formalism is that it allows to consistently use data at different energies, increasing the reach of the individual contributions by exploiting their synergy. For what concerns us here, semileptonic tau decays nicely complement both the traditional low-energy precision probes (Kaon and pion as well as nuclear beta decays) and the high-energy measurements (electroweak precision observables and LHC data).

The most general effective Lagrangian describing hadronic tau decays at dimension six is \cite{Cirigliano:2009wk}~\footnote{It can be obtained as the low-energy limit of the SM effective field theory (SMEFT) \cite{Grzadkowski:2010es}. Contributions from right-handed neutrinos vanish at the leading order in the $\epsilon_i$, to which we will stick.}
\begin{eqnarray}
 \mathcal{L}&=&-\frac{G_F V_{uD}}{\sqrt{2}}\Big[(1+\epsilon_L^\tau)\overline{\tau}\gamma_\mu(1-\gamma_5)\nu_\tau\cdot \overline{u}\gamma^\mu(1-\gamma_5)D+\epsilon_R^\tau\overline{\tau}\gamma_\mu(1-\gamma_5)\nu_\tau\cdot \overline{u}\gamma^\mu(1+\gamma_5)D\nonumber\\
 &&+\overline{\tau}(1-\gamma_5)\nu_\tau\cdot \overline{u}(\epsilon^\tau_S-\epsilon^\tau_P\gamma_5)D+\epsilon^\tau_T\overline{\tau}\sigma_{\mu\nu}(1-\gamma_5)\nu_\tau\cdot \overline{u}\sigma^{\mu\nu}(1-\gamma_5)D\Big]+\mathrm{h.c.}\,,
\end{eqnarray}
where $\sigma^{\mu\nu}=i[\gamma^\mu,\gamma^\nu]/2$ and $\epsilon_i$ $(i=L,R,S,P,T)$ are effective  couplings characterizing heavy new physics. The SM case is recovered for all $\epsilon_i=0$. Although we kept the lepton flavor ($\tau$) index in the $\epsilon_i$, we omitted the  corresponding quark flavor index ($D$). The $D=d$ and $D=s$ couplings would be the same according to minimal flavor violation (MFV) \cite{DAmbrosio:2002vsn}. We note that the product $G_F V_{ud}$ denotes its determination from superallowed nuclear Fermi $\beta$ decays (that includes, in fact, $\tilde{V}_{ud}^e$). As such~\cite{Gonzalez-Alonso:2016etj}, $G_F \tilde{V}_{uD}^e=G_F(1+\epsilon^e_L+\epsilon^e_R)V_{uD}$, which will introduce a dependence of our results on $\epsilon^e_L+\epsilon^e_R$. The new physics scale suppressing these interactions, can be related to the $\epsilon_i^\tau$ by $\Lambda\sim v/\sqrt{V_{uD}\epsilon_i}$, $v\sim246$ GeV.

This framework has been applied to study tau decays into the following meson systems: $\eta^{(\prime)}\pi^-$ \cite{Garces:2017jpz}, $\pi^-\pi^0$ \cite{Miranda:2018cpf}, $\pi^-(\pi^0/\eta)$ in combination with inclusive $\overline{u}\to d$ transitions \cite{Cirigliano:2018dyk}, $(K\pi)^-$ \cite{Rendon:2019awg} (with great focus on the CP asymmetry in $K_S\pi^-$ \cite{Cirigliano:2017tqn, Rendon:2019awg, Chen:2019vbr,Chen:2020uxi,Chen:2021udz}), $K\eta^{(\prime)}$ \cite{Gonzalez-Solis:2019lze} and a joint analysis of all one- and two-meson tau decays \cite{Gonzalez-Solis:2020jlh}, recently updated using improved radiative corrections for the $(\pi/K)^-$ modes \cite{Arroyo-Urena:2021nil}.

The $\eta^{(\prime)}\pi^-$ modes are very sensitive to non-SM scalar interactions, as the $\epsilon_S^\tau$ dependence is enhanced by a factor $\sqrt{s}/(m_d-m_u)$. Unfortunately, the big uncertainty in the corresponding FF of these yet undiscovered modes implies a large error on the obtained bounds. Still, the limits on $\epsilon_S^\tau$ \cite{Garces:2017jpz} from the non-observation of the $\eta\pi^-$ mode bind $\Lambda>3.5$ TeV at 90$\%$ C. L. (other limits given below correspond also to this confidence interval). The Belle-II discovery of this channel will allow to check these limits, and the measurement of its spectrum to increase the reach on new physics. Bounds are compatible, although slightly worse (and more uncertain) for the $\eta'\pi^-$ case.

The $\pi^-\pi^0 $ modes are quite sensitive to tensor interactions, $\epsilon_T^\tau$. Despite the parametric enhancement of $\epsilon_S^\tau$ contributions is the same as for $\eta^{(\prime)}\pi^-$, the dynamical suppression of the corresponding scalar FF  \cite{Descotes-Genon:2014tla}  spoils the prospects for binding it competitively. As in all other measured channels, a simultaneous fit of the (SM) dispersive FF parameters and of the $\epsilon_i$ to data is currently impossible, unfortunately. Notwithstanding, this could be achieved if Dalitz plot and angular distribution  measurements (these and other observables are covered extensively in the refs. quoted in this section) were done. We hope this (and other similar observations) motivates Belle-II analyses in this direction. The most stringent limit is obtained \cite{Miranda:2018cpf} from a fit to Belle data \cite{Belle:2008xpe} and binds $\Lambda>3.5$ TeV. More precise Belle-II measurements will probe higher energy scales.

CP violation has attracted a lot of attention towards the $\tau\to K_S\pi\nu_\tau$ decays. This was triggered by the BaBar measurement of the corresponding CP asymmetry
\begin{equation}
 A_{CP}^\tau=\frac{\Gamma(\tau^+\to\pi^+K_S\overline{\nu}_\tau)-\Gamma(\tau^-\to\pi^-K_S\nu_\tau)}{\Gamma(\tau^+\to\pi^+K_S\overline{\nu}_\tau)+\Gamma(\tau^-\to\pi^-K_S\nu_\tau)}\,,
\end{equation}
yielding $A_{CP}^\tau|_{\mathrm{exp}}={\color{red}-}3.6(2.3)(1.1)\times10^{-3}$ \cite{BaBar:2011pij}, which is $2.8\sigma$ away from the SM prediction, $A_{CP}^\tau|_{\mathrm{th}}=3.6(1)\times10^{-3}$ \cite{Grossman:2011zk}, that is determined by the minutely known neutral Kaon mixing \cite{ParticleDataGroup:2020ssz}. It must be noted that the corresponding Belle measurement \cite{Belle:2011sna} of a binned CP asymmetry was compatible with zero at the permille level in the four measured bins. Within the  SMEFT, it is impossible \cite{Cirigliano:2017tqn} that the anomaly found by BaBar can be explained as a result of heavy new physics. The corresponding contribution would be proportional to both the strong and the weak phase difference between the vector and tensor FFs. The former vanishes in the elastic region \cite{Watson:1952ji, Ecker:1988te, Cirigliano:2017tqn} and grows slowly, according to unitarity constraints, due to inelastic effects \cite{Cirigliano:2017tqn, Rendon:2019awg}. The latter must be, at most, $\sim\mathcal{O}(10^{-5})$ due to the experimental constraints on $D^0-\overline{D}^0$ mixing and the neutron electric dipole moment \cite{Cirigliano:2017tqn}. This restricts the heavy new physics  contribution to $A_{CP}^\tau$ to be $\leq 10^{-6}$ \cite{Cirigliano:2017tqn, Rendon:2019awg}, three orders of magnitude smaller than needed to explain the BaBar measurement. Turning to CP-conserving observables, the $\tau^-\to(K\pi)^-\nu_\tau$ decays have good sensitivity to both $\epsilon_{S,T}$. The corresponding bounds \cite{Rendon:2019awg} on them imply $\Lambda>3$ TeV. This limit agrees with those previously discussed for $\overline{u}\to d$ transitions and support the possible universality of the $\epsilon^\tau$ for $D=d,s$, \textit{i.e.} MFV. Again, we hope that Belle-II data on strangeness-changing tau decays will allow to be sensitive to higher new physics' scales.

The best limits set in exclusive hadronic tau decays are collected in Table \ref{tab}.

\begin{table}[h!]
    \centering
    \begin{tabular}{|c|c|c|}
    \hline
    Coefficient & Limit & Source\\
    \hline
     $\hat{\epsilon}^d_S$  & $-(2.4\pm5.3)\times10^{-3}$ & $\tau^-\to\eta\pi^-\nu_\tau$ \cite{Garces:2017jpz}\\
    $\hat{\epsilon}^d_T$     & $-\left(1.3^{+1.5}_{-2.2}\right)\times10^{-3}$ & $\tau^-\to\pi^0\pi^-\nu_\tau$ \cite{Miranda:2018cpf}\\
     $\hat{\epsilon}^s_S$   & $(1.3\pm0.9)\times10^{-2}$ & $\tau^-\to(K\pi)^-\nu_\tau$ \cite{Rendon:2019awg}\\
    $\hat{\epsilon}^s_T$     & $(0.7\pm1.0)\times10^{-2}$ & $\tau^-\to(K\pi)^-\nu_\tau$ \cite{Rendon:2019awg}\\
    \hline
    \end{tabular}
    \caption{Best limits on the $\hat\epsilon_i^D:=\epsilon^\tau_i/(1+\epsilon^e_L+\epsilon^e_R)$ coefficients, set from exclusive hadronic tau decays.}
    \label{tab}
\end{table}

Finally we turn to global fits of the $\epsilon_i$, which use different sets of data simultaneously. Combining $\tau^-\to\pi^-\nu_\tau$,  $\tau^-\to\pi^-\pi^0\nu_\tau$~\footnote{This channel was used via its impact on $a_\mu^{HVP,LO}$ \cite{Cirigliano:2001er, Cirigliano:2002pv}, see  \cite{Miranda:2020wdg} for an updated evaluation in the Resonance Chiral Lagrangian  framework \cite{Ecker:1988te,Ecker:1989yg,Cirigliano:2006hb,Kampf:2011ty}.}, $\tau^-\to\eta\pi^-\nu_\tau$ (with hadron input from \cite{Garces:2017jpz}) and the inclusive non-strange tau hadronic width, the following limits \cite{Cirigliano:2018dyk}
\begin{equation}\label{EqCiri}
\begin{pmatrix}
\epsilon^\tau_L-\epsilon^e_L+\epsilon^\tau_R-\epsilon^\tau_R\\
\epsilon^\tau_R\\
\epsilon^\tau_S\\
\epsilon^\tau_P\\
\epsilon^\tau_T
\end{pmatrix}
=
\begin{pmatrix}
1.0\pm1.1\\
0.2\pm1.3\\
-0.6\pm1.5\\
0.5\pm1.2\\
-0.04\pm0.46
\end{pmatrix}
\times10^{-2}\,,
\end{equation}
were set, corresponding to $\Lambda\gtrsim2$ TeV. Fig. 1 of Ref. \cite{Cirigliano:2018dyk} nicely shows that the addition of hadronic tau data to the electroweak precision observables \cite{Falkowski:2017pss} and LHC constraints allows to shrink the allowed contour in the plane defined by right-handed tau couplings vs. difference of left-handed couplings to taus and electrons by more than a factor two.

The separate analysis of the $\Delta S=0,1$ exclusive (one and two mesons) hadronic tau decays \cite{Gonzalez-Solis:2020jlh} cannot disentangle $\epsilon^\tau_{R,P}$. Limits on the $\epsilon_i$  
are, as in eq.(\ref{EqCiri}),  of $\mathcal{O}(10^{-2})$ for both $\Delta S=0,1$ and are given in the following:

\begin{equation}\label{DeltaS=0,1}
\begin{pmatrix}
\epsilon^\tau_L-\epsilon^e_L+\epsilon^\tau_R-\epsilon^\tau_R\\
\epsilon^\tau_R+\frac{m_\pi^2}{2M_\tau(m_u+m_d)}\epsilon^\tau_P\\
\epsilon^\tau_S\\
\epsilon^\tau_T
\end{pmatrix}
=
\begin{pmatrix}
0.5\pm2.2\\
0.3\pm1.2\\
-0.2\pm0.5\\
-0.1\pm1.3
\end{pmatrix}
\times10^{-2}\,,
\begin{pmatrix}
\epsilon^\tau_L-\epsilon^e_L+\epsilon^\tau_R-\epsilon^\tau_R\\
\epsilon^\tau_R+\frac{m_K^2}{2M_\tau(m_u+m_s)}\epsilon^\tau_P\\
\epsilon^\tau_S\\
\epsilon^\tau_T
\end{pmatrix}
=
\begin{pmatrix}
0.5\pm1.5\\
0.4\pm0.9\\
0.8\pm0.9\\
0.9\pm0.8
\end{pmatrix}
\times10^{-2}
\end{equation}
with the left(right) constraints corresponding to the $\Delta S=0(1)$ sectors. The compatibility of eqs. (\ref{EqCiri}) and (\ref{DeltaS=0,1}) confirms the robustness of these bounds. Assuming MFV, a joint fit of both $\Delta S=0,1$ sectors can be performed \cite{Gonzalez-Solis:2020jlh}. This untangles $\epsilon^\tau_{R,P}$ albeit at the price of a very big error on (and correlation between) them.
\section{Conclusion}
\label{sec:Concl}
Exclusive semileptonic tau decays remain to be a clean laboratory for increasing our knowledge on hadronization at low energies, where resonance properties (like their pole positions) can be determined very accurately. At the current level of precision, QCD-driven descriptions are necessary and benefit from experimental and lattice data by using dispersion relations and the known chiral and asymptotic limits. Such thorough understanding of these decays in the Standard Model enables a number of searches for new physics. Other talks in the conference have discussed lepton universality and CKM unitarity tests, searches for second class currents, CP and T violation studies, and analyses of $a_\mu^{HVP,LO}$ using tau data. Here focus was on the effective field theory analyses of these decays beyond the standard $W$ exchange, which bind  the corresponding new physics scale at a few TeV, competitively and complementary to Kaon, pion and nuclear beta decays as well as to electroweak precision observables or LHC data. Altogether, prospects for semileptonic tau decays, in light of Belle-II data, are bright.

\section*{Acknowledgements}
\hspace{0.5cm}I am indebted to all my collaborators in the topics here discussed. I thank the TAU2021 Organizing Committees' work, in difficult pandemic circumstances, and congratulate them for the very pleasant and fruitful conference. Even in this online format, I deeply missed Simon. Olya's passing away, so early, was terribly sad. Rest both in peace.

\paragraph{Funding information}
The support of Cátedras Marcos Moshinsky  (Fundación Marcos Moshinsky) is acknowledged.

\paragraph{Note added}Just before sending this manuscript, Ref. \cite{Cirigliano:2021yto} appeared. Although it is not covered in this contribution, it will be detailed in the proceedings of the next TAU conference.

\nolinenumbers


\begin{thebibliography}{99}

\bibitem{Davier:2005xq}
M.~Davier, A.~Hocker and Z.~Zhang,
``The Physics of Hadronic Tau Decays,''
Rev. Mod. Phys. \textbf{78} (2006), 1043-1109.
doi:10.1103/RevModPhys.78.1043.

\bibitem{Pich:2013lsa}
A.~Pich,
``Precision Tau Physics,''
Prog. Part. Nucl. Phys. \textbf{75} (2014), 41-85.
doi:10.1016/j.ppnp.2013.11.002.

\bibitem{Gonzalez-Solis:2021qzp}
S.~Gonz\`alez-Solís,
``Probes of non-standard interactions from exclusive hadronic tau decays,''
PoS \textbf{CHARM2020} (2021), 044.
doi:10.22323/1.385.0044.

\bibitem{ParticleDataGroup:2020ssz}
P.~A.~Zyla \textit{et al.} [Particle Data Group],
``Review of Particle Physics,''
PTEP \textbf{2020} (2020) no.8, 083C01.
doi:10.1093/ptep/ptaa104.

\bibitem{Kuhn:1992nz}
J.~H.~Kühn and E.~Mirkes,
``Structure functions in tau decays,''
Z. Phys. C \textbf{56} (1992), 661-672
[erratum: Z. Phys. C \textbf{67} (1995), 364].
doi:10.1007/BF01474741.

\bibitem{Weinberg:1978kz}
S.~Weinberg,
``Phenomenological Lagrangians,''
Physica A \textbf{96} (1979) no.1-2, 327-340.
doi:10.1016/0378-4371(79)90223-1.

\bibitem{Gasser:1983yg}
J.~Gasser and H.~Leutwyler,
``Chiral Perturbation Theory to One Loop,''
Annals Phys. \textbf{158} (1984), 142.
doi:10.1016/0003-4916(84)90242-2.

\bibitem{Gasser:1984gg}
J.~Gasser and H.~Leutwyler,
``Chiral Perturbation Theory: Expansions in the Mass of the Strange Quark,''
Nucl. Phys. B \textbf{250} (1985), 465-516.
doi:10.1016/0550-3213(85)90492-4.

\bibitem{Lepage:1980fj}
G.~P.~Lepage and S.~J.~Brodsky,
``Exclusive Processes in Perturbative Quantum Chromodynamics,''
Phys. Rev. D \textbf{22} (1980), 2157.
doi:10.1103/PhysRevD.22.2157.

\bibitem{FlavourLatticeAveragingGroup:2019iem}
S.~Aoki \textit{et al.} [Flavour Lattice Averaging Group],
``FLAG Review 2019: Flavour Lattice Averaging Group (FLAG),''
Eur. Phys. J. C \textbf{80} (2020) no.2, 113.
doi:10.1140/epjc/s10052-019-7354-7.

\bibitem{Guo:2010dv}
Z.~H.~Guo and P.~Roig,
``One meson radiative tau decays,''
Phys. Rev. D \textbf{82} (2010), 113016.
doi:10.1103/PhysRevD.82.113016.

\bibitem{Guevara:2013wwa}
A.~Guevara, G.~L\'opez Castro and P.~Roig,
``Weak radiative pion vertex in $\tau^- \to \pi^-\nu_\tau \ell^+ \ell^-$ decays,''
Phys. Rev. D \textbf{88} (2013) no.3, 033007.
doi:10.1103/PhysRevD.88.033007
, and ``Improved description of di-lepton production in $\tau^-\to\nu_\tau P^-$ decays,''
[arXiv:2111.09994 [hep-ph]].

\bibitem{Arroyo-Urena:2021nil}
M.~A.~Arroyo-Ure\~na, G.~Hern\'andez-Tom\'e, G.~L\'opez-Castro, P.~Roig and I.~Rosell,
``Radiative corrections to $\tau \to \pi (K) \nu_\tau [\gamma]$: a reliable new physics test,''
[arXiv:2107.04603 [hep-ph]], and  
``One-loop determination of $\tau \to \pi (K) \nu_{\tau}[\gamma]$ branching ratios and new physics tests,''
[arXiv:2112.01859 [hep-ph]].

\bibitem{OPAL:2000fde}
G.~Abbiendi \textit{et al.} [OPAL],
``A Study of one prong tau decays with a charged kaon,''
Eur. Phys. J. C \textbf{19} (2001), 653-665.
doi:10.1007/s100520100632.

\bibitem{ALEPH:2005qgp}
S.~Schael \textit{et al.} [ALEPH],
``Branching ratios and spectral functions of tau decays: Final ALEPH measurements and physics implications,''
Phys. Rept. \textbf{421} (2005), 191-284.
doi:10.1016/j.physrep.2005.06.007.

\bibitem{BaBar:2009lyd}
B.~Aubert \textit{et al.} [BaBar],
``Measurements of Charged Current Lepton Universality and $|V_{us}|$ using Tau Lepton Decays to $e^- \bar{\nu}_e \nu_\tau$, $\mu^- \bar{\nu}_\mu \nu_\tau$, $\pi^- \nu_\tau$, and $K^- \nu_\tau$,''
Phys. Rev. Lett. \textbf{105} (2010), 051602.
doi:10.1103/PhysRevLett.105.051602.

\bibitem{Belle:2019bpr}
Y.~Jin \textit{et al.} [Belle],
``Observation of $ \tau^- \rightarrow \pi^- \nu_{\tau} e^+ e^- $ and search for $\tau^- \rightarrow \pi^- \nu_{\tau} \mu^+ \mu^-$,''
Phys. Rev. D \textbf{100} (2019) no.7, 071101.
doi:10.1103/PhysRevD.100.071101.

\bibitem{Gasser:1984ux}
J.~Gasser and H.~Leutwyler,
``Low-Energy Expansion of Meson Form-Factors,''
Nucl. Phys. B \textbf{250} (1985), 517-538.
doi:10.1016/0550-3213(85)90493-6.

\bibitem{Oller:2000ug}
J.~A.~Oller, E.~Oset and J.~E.~Palomar,
``Pion and kaon vector form-factors,''
Phys. Rev. D \textbf{63} (2001), 114009.
doi:10.1103/PhysRevD.63.114009.

\bibitem{Gonzalez-Solis:2019iod}
S.~Gonz\`alez-Solís and P.~Roig,
``A dispersive analysis of the pion vector form factor and $\tau ^{-}\rightarrow K^{-}K_{S}\nu _{\tau }$ decay,''
Eur. Phys. J. C \textbf{79} (2019) no.5, 436.
doi:10.1140/epjc/s10052-019-6943-9.

\bibitem{Pich:2001pj}
A.~Pich and J.~Portolés,
``The Vector form-factor of the pion from unitarity and analyticity: A Model independent approach,''
Phys. Rev. D \textbf{63} (2001), 093005.
doi:10.1103/PhysRevD.63.093005.

\bibitem{GomezDumm:2013sib}
D.~G\'omez Dumm and P.~Roig,
``Dispersive representation of the pion vector form factor in $\tau\to\pi\pi\nu_\tau$ decays,''
Eur. Phys. J. C \textbf{73} (2013) no.8, 2528.
doi:10.1140/epjc/s10052-013-2528-1.

\bibitem{Watson:1952ji}
K.~M.~Watson,
``The Effect of final state interactions on reaction cross-sections,''
Phys. Rev. \textbf{88} (1952), 1163-1171.
doi:10.1103/PhysRev.88.1163.

\bibitem{Colangelo:2001df}
G.~Colangelo, J.~Gasser and H.~Leutwyler,
``$\pi \pi$ scattering,''
Nucl. Phys. B \textbf{603} (2001), 125-179.
doi:10.1016/S0550-3213(01)00147-X.

\bibitem{Garcia-Martin:2011iqs}
R.~García-Martín, R.~Kaminski, J.~R.~Peláez, J.~Ruiz de Elvira and F.~J.~Ynduráin,
``The Pion-pion scattering amplitude. IV: Improved analysis with once subtracted Roy-like equations up to 1100 MeV,''
Phys. Rev. D \textbf{83} (2011), 074004.
doi:10.1103/PhysRevD.83.074004.

\bibitem{Caprini:2011ky}
I.~Caprini, G.~Colangelo and H.~Leutwyler,
``Regge analysis of the pi pi scattering amplitude,''
Eur. Phys. J. C \textbf{72} (2012), 1860.
doi:10.1140/epjc/s10052-012-1860-1.

\bibitem{OPAL:1998rrm}
K.~Ackerstaff \textit{et al.} [OPAL],
``Measurement of the strong coupling constant alpha(s) and the vector and axial vector spectral functions in hadronic tau decays,''
Eur. Phys. J. C \textbf{7} (1999), 571-593.
doi:10.1007/s100529901061.

\bibitem{CLEO:1999dln}
S.~Anderson \textit{et al.} [CLEO],
``Hadronic structure in the decay tau-> pi- pi0 neutrino(tau),''
Phys. Rev. D \textbf{61} (2000), 112002.
doi:10.1103/PhysRevD.61.112002.

\bibitem{Belle:2008xpe}
M.~Fujikawa \textit{et al.} [Belle],
``High-Statistics Study of the tau- -> pi- pi0 nu(tau) Decay,''
Phys. Rev. D \textbf{78} (2008), 072006.
doi:10.1103/PhysRevD.78.072006.

\bibitem{BaBar:2018qry}
J.~P.~Lees \textit{et al.} [BaBar],
``Measurement of the spectral function for the $\tau^-\to K^-K_S\nu_{\tau}$ decay,''
Phys. Rev. D \textbf{98} (2018) no.3, 032010.
doi:10.1103/PhysRevD.98.032010.

\bibitem{Cirigliano:2001er}
V.~Cirigliano, G.~Ecker and H.~Neufeld,
``Isospin violation and the magnetic moment of the muon,''
Phys. Lett. B \textbf{513} (2001), 361-370.
doi:10.1016/S0370-2693(01)00764-X.

\bibitem{Cirigliano:2002pv}
V.~Cirigliano, G.~Ecker and H.~Neufeld,
``Radiative tau decay and the magnetic moment of the muon,''
JHEP \textbf{08} (2002), 002.
doi:10.1088/1126-6708/2002/08/002.

\bibitem{Flores-Baez:2006yiq}
F.~Flores-Báez, A.~Flores-Tlalpa, G.~López Castro and G.~Toledo Sánchez,
``Long-distance radiative corrections to the di-pion tau lepton decay,''
Phys. Rev. D \textbf{74} (2006), 071301.
doi:10.1103/PhysRevD.74.071301.

\bibitem{Miranda:2020wdg}
J.~A.~Miranda and P.~Roig,
``New $\tau$-based evaluation of the hadronic contribution to the vacuum polarization piece of the muon anomalous magnetic moment,''
Phys. Rev. D \textbf{102} (2020), 114017.
doi:10.1103/PhysRevD.102.114017.

\bibitem{GutierrezSantiago:2020bhy}
J.~L.~Guti\'errez Santiago, G.~L\'opez Castro and P.~Roig,
``Lepton-pair production in dipion $\tau$ lepton decays,''
Phys. Rev. D \textbf{103} (2021) no.1, 014027 .
doi:10.1103/PhysRevD.103.014027.

\bibitem{Aoyama:2020ynm}
T.~Aoyama, 
\textit{et al.}
``The anomalous magnetic moment of the muon in the Standard Model,''
Phys. Rept. \textbf{887} (2020), 1-166.
doi:10.1016/j.physrep.2020.07.006.

\bibitem{Jamin:2000wn}
M.~Jamin, J.~A.~Oller and A.~Pich,
``S wave K pi scattering in chiral perturbation theory with resonances,''
Nucl. Phys. B \textbf{587} (2000), 331-362.
doi:10.1016/S0550-3213(00)00479-X.

\bibitem{Jamin:2001zq}
M.~Jamin, J.~A.~Oller and A.~Pich,
``Strangeness changing scalar form-factors,''
Nucl. Phys. B \textbf{622} (2002), 279-308.
doi:10.1016/S0550-3213(01)00605-8.

\bibitem{Moussallam:2007qc}
B.~Moussallam,
``Analyticity constraints on the strangeness changing vector current and applications to tau -> K pi nu(tau), tau -> K pi pi nu(tau),''
Eur. Phys. J. C \textbf{53} (2008), 401-412.
doi:10.1140/epjc/s10052-007-0464-7.

\bibitem{Boito:2008fq}
D.~R.~Boito, R.~Escribano and M.~Jamin,
``K pi vector form-factor, dispersive constraints and tau -> nu(tau) K pi decays,''
Eur. Phys. J. C \textbf{59} (2009), 821-829 .
doi:10.1140/epjc/s10052-008-0834-9.

\bibitem{Boito:2010me}
D.~R.~Boito, R.~Escribano and M.~Jamin,
``K $\pi$ vector form factor constrained by $\tau -> K\ pi \nu_\tau$ and $K_{l3}$ decays,''
JHEP \textbf{09} (2010), 031.
doi:10.1007/JHEP09(2010)031.

\bibitem{Antonelli:2013usa}
M.~Antonelli, V.~Cirigliano, A.~Lusiani and E.~Passemar,
``Predicting the $\tau$ strange branching ratios and implications for $V_{us}$,''
JHEP \textbf{10} (2013), 070.
doi:10.1007/JHEP10(2013)070.

\bibitem{Bernard:2013jxa}
V.~Bernard,
``First determination of $f_+(0) |V_{us}|$ from a combined analysis of $\tau\to K\pi \nu_\tau$ decay and $\pi K$ scattering with constraints from $K_{\ell3}$ decays,''
JHEP \textbf{06} (2014), 082.
doi:10.1007/JHEP06(2014)082.

\bibitem{Escribano:2013bca}
R.~Escribano, S.~Gonzàlez-Solís and P.~Roig,
``$\tau^-\to K^-\eta^{(\prime)}\nu_\tau$ decays in Chiral Perturbation Theory with Resonances,''
JHEP \textbf{10} (2013), 039.
doi:10.1007/JHEP10(2013)039.

\bibitem{Belle:2007goc}
D.~Epifanov \textit{et al.} [Belle],
``Study of tau- -> K(S) pi- nu(tau) decay at Belle,''
Phys. Lett. B \textbf{654} (2007), 65-73.
doi:10.1016/j.physletb.2007.08.045.

\bibitem{Belle:2008jjb}
K.~Inami \textit{et al.} [Belle],
``Precise measurement of hadronic tau-decays with an eta meson,''
Phys. Lett. B \textbf{672} (2009), 209-218.
doi:10.1016/j.physletb.2009.01.047.

\bibitem{Escribano:2014joa}
R.~Escribano, S.~Gonz\'alez-Solís, M.~Jamin and P.~Roig,
``Combined analysis of the decays $\tau^{-} \to K_{S} \pi^{-} \nu_{\tau}$ and $\tau^{-} \to K^{-} \eta\nu_{\tau}$,''
JHEP \textbf{09} (2014), 042.
doi:10.1007/JHEP09(2014)042.
[arXiv:1407.6590 [hep-ph]].

\bibitem{BaBar:2007yir}
B.~Aubert \textit{et al.} [BaBar],
``Measurement of the $\tau^{-} \to K^{-} \pi^0 \nu_{tau}$ branching fraction,''
Phys. Rev. D \textbf{76} (2007), 051104.
doi:10.1103/PhysRevD.76.051104.

\bibitem{BaBar:2012zfq}
J.~P.~Lees \textit{et al.} [BaBar],
``Study of high-multiplicity 3-prong and 5-prong tau decays at BABAR,''
Phys. Rev. D \textbf{86} (2012), 092010.
doi:10.1103/PhysRevD.86.092010.

\bibitem{BaBar:2010bul}
P.~del Amo Sanchez \textit{et al.} [BaBar],
``Studies of tau-> eta K-nu and tau-> eta pi- nu(tau) at BaBar and a search for a second-class current,''
Phys. Rev. D \textbf{83} (2011), 032002.
doi:10.1103/PhysRevD.83.032002.

\bibitem{Guevara:2016trs}
A.~Guevara, G.~L\'opez-Castro and P.~Roig,
``$\tau^-\to\eta^{(\prime)}\pi^-\nu_\tau \gamma$ decays as backgrounds in the search for second class currents,''
Phys. Rev. D \textbf{95} (2017) no.5, 054015.
doi:10.1103/PhysRevD.95.054015.

\bibitem{Hernandez-Tome:2017pdc}
G.~Hern\'andez-Tom\'e, G.~L\'opez Castro and P.~Roig,
``G-parity breaking in $\tau^- \to \eta^{(\prime)} \pi^- \nu_{\tau}$ decays induced by the $\eta^{(\prime)}\gamma\gamma$ form factor,''
Phys. Rev. D \textbf{96} (2017) no.5, 053003.
doi:10.1103/PhysRevD.96.053003.

\bibitem{Belle-II:2018jsg}
E.~Kou \textit{et al.} [Belle-II],
``The Belle II Physics Book,''
PTEP \textbf{2019} (2019) no.12, 123C01
[erratum: PTEP \textbf{2020} (2020) no.2, 029201].
doi:10.1093/ptep/ptz106.

\bibitem{Descotes-Genon:2014tla}
S.~Descotes-Genon and B.~Moussallam,
``Analyticity of $\eta \pi $ isospin-violating form factors and the $\tau \rightarrow \eta \pi \nu $ second-class decay,''
Eur. Phys. J. C \textbf{74} (2014), 2946.
doi:10.1140/epjc/s10052-014-2946-8.

\bibitem{Escribano:2016ntp}
R.~Escribano, S.~Gonzàlez-Solís and P.~Roig,
``Predictions on the second-class current decays $\tau^{-}\to\pi^{-}\eta^{(\prime)}\nu_{\tau}$,''
Phys. Rev. D \textbf{94} (2016) no.3, 034008.
doi:10.1103/PhysRevD.94.034008.

\bibitem{Guo:2012yt}
Z.~H.~Guo, J.~A.~Oller and J.~Ruiz de Elvira,
``Chiral dynamics in form factors, spectral-function sum rules, meson-meson scattering and semi-local duality,''
Phys. Rev. D \textbf{86} (2012), 054006.
doi:10.1103/PhysRevD.86.054006.

\bibitem{Dudek:2016cru}
J.~J.~Dudek \textit{et al.} [Hadron Spectrum],
``An $a_0$ resonance in strongly coupled $\pi \eta$, $K\overline{K}$ scattering from lattice QCD,''
Phys. Rev. D \textbf{93} (2016) no.9, 094506.
doi:10.1103/PhysRevD.93.094506.

\bibitem{Guo:2016zep}
Z.~H.~Guo, L.~Liu, U.~G.~Mei\ss{}ner, J.~A.~Oller and A.~Rusetsky,
``Chiral study of the $a_0(980)$ resonance and $\pi\eta$ scattering phase shifts in light of a recent lattice simulation,''
Phys. Rev. D \textbf{95} (2017) no.5, 054004.
doi:10.1103/PhysRevD.95.054004.

\bibitem{Colangelo:1996hs}
G.~Colangelo, M.~Finkemeier and R.~Urech,
``Tau decays and chiral perturbation theory,''
Phys. Rev. D \textbf{54} (1996), 4403-4418.
doi:10.1103/PhysRevD.54.4403.

\bibitem{GomezDumm:2003ku}
D.~Gómez Dumm, A.~Pich and J.~Portolés,
``tau -> pi pi pi nu(tau) decays in the resonance effective theory,''
Phys. Rev. D \textbf{69} (2004), 073002.
doi:10.1103/PhysRevD.69.073002.

\bibitem{Dumm:2009va}
D.~G.~Dumm, P.~Roig, A.~Pich and J.~Portolés,
``tau -> pi pi pi nu(tau) decays and the a(1)(1260) off-shell width revisited,''
Phys. Lett. B \textbf{685} (2010), 158-164.
doi:10.1016/j.physletb.2010.01.059.

\bibitem{Dumm:2009kj}
D.~G.~Dumm, P.~Roig, A.~Pich and J.~Portolés,
``Hadron structure in tau -> KK pi nu (tau) decays,''
Phys. Rev. D \textbf{81} (2010), 034031.
doi:10.1103/PhysRevD.81.034031.

\bibitem{GomezDumm:2012dpx}
D.~Gómez Dumm and P.~Roig,
``Resonance Chiral Lagrangian analysis of $\tau^- \to \eta^{(\prime)} \pi^- \pi^0 \nu_\tau$ decays,''
Phys. Rev. D \textbf{86} (2012), 076009.
doi:10.1103/PhysRevD.86.076009.

\bibitem{Shekhovtsova:2012ra}
O.~Shekhovtsova, T.~Przedzinski, P.~Roig and Z.~Was,
``Resonance chiral Lagrangian currents and $\tau$ decay Monte Carlo,''
Phys. Rev. D \textbf{86} (2012), 113008.
doi:10.1103/PhysRevD.86.113008.

\bibitem{Nugent:2013hxa}
I.~M.~Nugent, T.~Przedzinski, P.~Roig, O.~Shekhovtsova and Z.~Was,
``Resonance chiral Lagrangian currents and experimental data for $\tau^-\to\pi^{-}\pi^{-}\pi^{+}\nu_{\tau}$,''
Phys. Rev. D \textbf{88} (2013), 093012.
doi:10.1103/PhysRevD.88.093012.

\bibitem{Jadach:1990mz}
S.~Jadach, J.~H.~Kühn and Z.~Was,
``TAUOLA: A Library of Monte Carlo programs to simulate decays of polarized tau leptons,''
Comput. Phys. Commun. \textbf{64} (1990), 275-299.
doi:10.1016/0010-4655(91)90038-M.

\bibitem{Jadach:1993hs}
S.~Jadach, Z.~Was, R.~Decker and J.~H.~Kühn,
``The tau decay library TAUOLA: Version 2.4,''
Comput. Phys. Commun. \textbf{76} (1993), 361-380.
doi:10.1016/0010-4655(93)90061-G.

\bibitem{Guo:2008sh}
Z.~H.~Guo,
``Study of tau-> V P nu(tau) in the framework of resonance chiral theory,''
Phys. Rev. D \textbf{78} (2008), 033004.
doi:10.1103/PhysRevD.78.033004.

\bibitem{JPAC:2018zwp}
M.~Mikhasenko \textit{et al.} [JPAC],
``Pole position of the $a_1(1260)$ from $\tau$-decay,''
Phys. Rev. D \textbf{98} (2018) no.9, 096021.
doi:10.1103/PhysRevD.98.096021.

\bibitem{ALEPH:1998rgl}
R.~Barate \textit{et al.} [ALEPH],
``Measurement of the spectral functions of axial - vector hadronic tau decays and determination of alpha(S)(M**2(tau)),''
Eur. Phys. J. C \textbf{4} (1998), 409-431.
doi:10.1007/s100520050217.

\bibitem{ALEPH:1999jxs}
R.~Barate \textit{et al.} [ALEPH],
``One prong tau decays with kaons,''
Eur. Phys. J. C \textbf{10} (1999), 1-18.
doi:10.1007/s100529900146.

\bibitem{ALEPH:1999uux}
R.~Barate \textit{et al.} [ALEPH],
``Study of tau decays involving kaons, spectral functions and determination of the strange quark mass,''
Eur. Phys. J. C \textbf{11} (1999), 599-618.
doi:10.1007/s100520050659.

\bibitem{CLEO:2004hrb}
T.~E.~Coan \textit{et al.} [CLEO],
``Wess-Zumino current and the structure of the decay tau-> K- K+ pi- nu(tau),''
Phys. Rev. Lett. \textbf{92} (2004), 232001.
doi:10.1103/PhysRevLett.92.232001.

\bibitem{Cirigliano:2009wk}
V.~Cirigliano, J.~Jenkins and M.~Gonzalez-Alonso,
Nucl. Phys. B \textbf{830} (2010), 95-115.
doi:10.1016/j.nuclphysb.2009.12.020.

\bibitem{Grzadkowski:2010es}
B.~Grzadkowski, M.~Iskrzynski, M.~Misiak and J.~Rosiek,
``Dimension-Six Terms in the Standard Model Lagrangian,''
JHEP \textbf{10} (2010), 085.
doi:10.1007/JHEP10(2010)085.

\bibitem{DAmbrosio:2002vsn}
G.~D'Ambrosio, G.~F.~Giudice, G.~Isidori and A.~Strumia,
``Minimal flavor violation: An Effective field theory approach,''
Nucl. Phys. B \textbf{645} (2002), 155-187.
doi:10.1016/S0550-3213(02)00836-2.

\bibitem{Gonzalez-Alonso:2016etj}
M.~Gonz\'alez-Alonso and J.~Martín Camalich,
``Global Effective-Field-Theory analysis of New-Physics effects in (semi)leptonic kaon decays,''
JHEP \textbf{12} (2016), 052.
doi:10.1007/JHEP12(2016)052.

\bibitem{Garces:2017jpz}
E.~A.~Garc\'es, M.~Hern\'andez Villanueva, G.~L\'opez Castro and P.~Roig,
``Effective-field theory analysis of the $\tau^- \to \eta^{(\prime)} \pi^- \nu_\tau$ decays,''
JHEP \textbf{12} (2017), 027.
doi:10.1007/JHEP12(2017)027.

\bibitem{Miranda:2018cpf}
J.~A.~Miranda and P.~Roig,
``Effective-field theory analysis of the $\tau^-\to \pi^-\pi^0\nu_\tau$ decays,''
JHEP \textbf{11} (2018), 038.
doi:10.1007/JHEP11(2018)038.

\bibitem{Cirigliano:2018dyk}
V.~Cirigliano, A.~Falkowski, M.~Gonz\'alez-Alonso and A.~Rodr\'\i{}guez-S\'anchez,
``Hadronic $\tau$ Decays as New Physics Probes in the LHC Era,''
Phys. Rev. Lett. \textbf{122} (2019) no.22, 221801.
doi:10.1103/PhysRevLett.122.221801.

\bibitem{Rendon:2019awg}
J.~Rend\'on, P.~Roig and G.~Toledo S\'anchez,
``Effective-field theory analysis of the $\tau^{-}\rightarrow (K \pi)^{-}\nu_{\tau}$ decays,''
Phys. Rev. D \textbf{99} (2019) no.9, 093005.
doi:10.1103/PhysRevD.99.093005.

\bibitem{Cirigliano:2017tqn}
V.~Cirigliano, A.~Crivellin and M.~Hoferichter,
``No-go theorem for nonstandard explanations of the $\tau\to K_S\pi\nu_\tau$ CP asymmetry,''
Phys. Rev. Lett. \textbf{120} (2018) no.14, 141803.
doi:10.1103/PhysRevLett.120.141803.

\bibitem{Chen:2019vbr}
F.~Z.~Chen, X.~Q.~Li, Y.~D.~Yang and X.~Zhang,
``CP asymmetry in $\tau\to K_S\pi\nu_\tau$ decays within the Standard Model and beyond,''
Phys. Rev. D \textbf{100} (2019) no.11, 113006.
doi:10.1103/PhysRevD.100.113006.

\bibitem{Chen:2020uxi}
F.~Z.~Chen, X.~Q.~Li and Y.~D.~Yang,
``$CP$ asymmetry in the angular distribution of $\tau\to K_S\pi\nu_\tau$ decays,''
JHEP \textbf{05} (2020), 151.
doi:10.1007/JHEP05(2020)151.

\bibitem{Chen:2021udz}
F.~Z.~Chen, X.~Q.~Li, S.~C.~Peng, Y.~D.~Yang and H.~H.~Zhang,
``$CP$ asymmetry in the angular distributions of $\tau\to K_S\pi\nu_\tau$ decays -- II: general effective field theory analysis,''
[arXiv:2107.12310 [hep-ph]].

\bibitem{Gonzalez-Solis:2019lze}
S.~Gonz\`alez-Sol\'\i{}s, A.~Miranda, J.~Rend\'on and P.~Roig,
``Effective-field theory analysis of the $\tau^{-}\to K^{-}(\eta^{(\prime)},K^{0}) \nu_{\tau}$ decays,''
Phys. Rev. D \textbf{101} (2020) no.3, 034010.
doi:10.1103/PhysRevD.101.034010.

\bibitem{Gonzalez-Solis:2020jlh}
S.~Gonz\`alez-Sol\'\i{}s, A.~Miranda, J.~Rend\'on and P.~Roig,
``Exclusive hadronic tau decays as probes of non-SM interactions,''
Phys. Lett. B \textbf{804} (2020), 135371.
doi:10.1016/j.physletb.2020.135371.

\bibitem{BaBar:2011pij}
J.~P.~Lees \textit{et al.} [BaBar],
``Search for CP Violation in the Decay $\tau^- -> \pi^- K^0_S (>= 0 \pi^0) \nu_tau$,''
Phys. Rev. D \textbf{85} (2012), 031102
[erratum: Phys. Rev. D \textbf{85} (2012), 099904].
doi:10.1103/PhysRevD.85.031102.

\bibitem{Grossman:2011zk}
Y.~Grossman and Y.~Nir,
``CP Violation in $\tau^\pm \to \pi^\pm K_S\nu$ and $D^\pm \to \pi^\pm K_S$: The Importance of $K_S - K_L$ Interference,''
JHEP \textbf{04} (2012), 002.
doi:10.1007/JHEP04(2012)002.

\bibitem{Belle:2011sna}
M.~Bischofberger \textit{et al.} [Belle],
``Search for CP violation in $\tau \to K^0_S \pi \nu_\tau$ decays at Belle,''
Phys. Rev. Lett. \textbf{107} (2011), 131801.
doi:10.1103/PhysRevLett.107.131801.

\bibitem{Ecker:1988te}
G.~Ecker, J.~Gasser, A.~Pich and E.~de Rafael,
``The Role of Resonances in Chiral Perturbation Theory,''
Nucl. Phys. B \textbf{321} (1989), 311-342.
doi:10.1016/0550-3213(89)90346-5.

\bibitem{Ecker:1989yg}
G.~Ecker, J.~Gasser, H.~Leutwyler, A.~Pich and E.~de Rafael,
``Chiral Lagrangians for Massive Spin 1 Fields,''
Phys. Lett. B \textbf{223} (1989), 425-432.
doi:10.1016/0370-2693(89)91627-4.

\bibitem{Cirigliano:2006hb}
V.~Cirigliano, G.~Ecker, M.~Eidemüller, R.~Kaiser, A.~Pich and J.~Portolés,
``Towards a consistent estimate of the chiral low-energy constants,''
Nucl. Phys. B \textbf{753} (2006), 139-177.
doi:10.1016/j.nuclphysb.2006.07.010.

\bibitem{Kampf:2011ty}
K.~Kampf and J.~Novotny,
``Resonance saturation in the odd-intrinsic parity sector of low-energy QCD,''
Phys. Rev. D \textbf{84} (2011), 014036.
doi:10.1103/PhysRevD.84.014036.

\bibitem{Falkowski:2017pss}
A.~Falkowski, M.~Gonz\'alez-Alonso and K.~Mimouni,
``Compilation of low-energy constraints on 4-fermion operators in the SMEFT,''
JHEP \textbf{08} (2017), 123.
doi:10.1007/JHEP08(2017)123.

\bibitem{Cirigliano:2021yto}
V.~Cirigliano, D.~D\'\i{}az-Calder\'on, A.~Falkowski, M.~Gonz\'alez-Alonso and A.~Rodr\'\i{}guez-S\'anchez,
``Semileptonic tau decays beyond the Standard Model,''
[arXiv:2112.02087 [hep-ph]].

\end{thebibliography}
\end{document}